# Electric field control of domain wall logic in piezoelectric/ferromagnetic nanodevices

Na Lei[1,2], Thibaut Devolder[1,2], Guillaume Agnus[1,2], Pascal Aubert[1,2], Laurent Daniel[3], Joo-Von Kim[1,2], Weisheng Zhao[1,2], Claude Chappert[1,2], Dafiné Ravelosona[1,2] and Philippe Lecoeur[1,2]

[1] Institut d'Electronique Fondamentale, Université Paris-Sud, 91405 Orsay, France
[2] UMR 8622, CNRS, 91405 Orsay, France
[3] Laboratoire de Génie Electrique de Paris, CNRS, UMR8507/SUPELEC/UPMC/Univ Paris-Sud, 91192 Gif-sur-Yvette, France.

**Power dissipation is one of the most important factors limiting the future miniaturisation of integrated circuits[1]. The capability of controlling magnetic states with a low voltage through magnetoelectric coupling in magnetostrictive/piezoelectric systems may pave the way toward ultra low-power electronics[2-4]. Although the former effect has been demonstrated in several multiferroic heterostructures, the incorporation of such complex geometries into practical magnetic memory and logic nanodevices has been lacking. Here, we demonstrate the room temperature control of a domain wall gate with an electric field in a nanowire consisting of a laterally polarized piezoelectric bar inducing a giant strain in a ferromagnetic spin-valve. We propose to use such novel domain wall gate as an elementary brick to generate a complete set of boolean logic functions or stabilize domain walls in high density memory applications.**

The prospect of controlling local magnetisation using electric fields for ultra low power spintronics has resulted in a new research direction that involves multiferroic and magnetoelectric materials[5-10]. However, existing approaches involving natural multiferroics are limited by weak ferromagnetism at room temperature[10] or are effective only at low temperatures[12,13]. One promising avenue to circumvent such issues concerns the use of strain-mediated coupling in piezoelectric/magnetostrictive bilayer structures[14-20]. In this structure, the applied voltage across the thickness generates a uniaxial strain in the piezoelectric layer that is transferred to the magnetostrictive nanomagnet, resulting in a change in its magnetic properties. It has been shown theoretically that this can be the basis of ultra low power computing and signal processing towards possible applications in the field of power harvesting[2,4]. Within this approach, first investigation of individual switching events under applied voltages has been reported in thin film FeGa/BaTiO$_3$ bilayer structures[21]. However, the difficulty to integrate such complex heterostructures geometries into electrically controlled nanoscale devices has precluded the demonstration of local manipulation of single nanomagnet that would be an essential step towards the magnetoelectric control of spintronic devices.

Domain wall propagating in magnetic nanowires are the devices of choice for ultra high density magnetic memory[22] and spin logic elements[23]. In this work, we demonstrate that domain wall motion can be electrically controlled at room temperature using strain mediated magnetoelectric coupling in piezoelectric/ferromagnetic nanostructure. We first describe a novel device geometry that allows for an efficient transfer of a strong local strain to the ferromagnetic stripe. Using conventional 3d ferromagnetic spin valves together with the commonly used piezoelectric material PZT, our experiments show that the energy barrier for domain wall motion can be double under reasonable electric field. Beyond this demonstration of an integrated all electrical magnetoelectronic device compatible with low power consumption high speed and high density, we propose to use such domain wall gate to generate Boolean logic functions or stabilize domain walls in memory applications..

A straightforward means of obtaining high magnetoelectric coupling involves encapsulating the magnetic nanodevice inside a piezoelectrical environment[3]. However, this approach faces two main difficulties. First, ceramic piezoelectric layers like PZT require high temperature deposition or post annealing (>400°C) under severe oxidation conditions[24], which inevitably degrade the magnetic properties. Second, the clamping effect from the substrate limits the crystal stress in any solid-state device. In order to avoid these issues, we have focused on a novel approach based on the lateral geometry seen in Figure 1, which is favourable to manipulate efficiently magnetic domain wall in nanowires (see Supplementary Information).

The sides of the piezoelectric bar are free to move but its bottom surface is clamped by the substrate, which does not allow for any global longitudinal compression. Applying the voltage re-sults in the electrical polarisation being rotated away from the growth direction[25], which induces a strain $\varepsilon_{xx} > 0$ corresponding to an elongation in the bar width, particularly near the top surface where the magnetic wire is located. In addition, this geometry leads to a large accumulated strain $\varepsilon_{yy}$ at both longitudinal ends of electrodes. Both strains along *x* and *y* contribute efficiently to pin domain walls irrespective of the sign of the magnetostriction parameter. The final device structure is shown in Figure 2a.

In addition to the strain being transferred efficiently to the ferromagnetic nanowire, it is necessary to ensure that magnetisation reversal, involving the easy propagation of a single domain wall, can be detected by a magnetotransport scheme compatible with integrated all electrical devices[26]. For this purpose, we used magnetic wires based on in-plane magnetised CoFeB/Cu/Co spin valves that exhibit a GMR effect. A proper choice of the lateral width of the stripe (700 nm) forces the wall to adopt a transverse magnetisation, with the sole degree of freedom being its position along the wire, and the sole relevant field component to move it being that parallel to the wire (y) axis (see supplementary material). The final device is shown in Figure 2a and 2b.

Transport measurements were first performed on the device without any applied voltage, as shown in Figure 2c. The typical GMR ratio is about 0.4%, which is sufficiently high to detect domain wall motion with good precision in the free layer. The presence of an asymmetric high resistance state results from the pinned layer (PL) magnetisation that starts to reverse before the free layer (FL) magnetisation is fully reversed. The magnetisation reversal of the PL is shifted to positive fields, due to the presence of an exchange bias field $\mu_0 H_{ex} = 15$ mT resulting from a coupling with the antiferromagnetic IrMn layer. For the FL, a slight shift towards positive fields (around 0.6 mT) can also be observed, which originates from an interlayer coupling between the two ferromagnetic layers[27]. Angular dependence of the coercive field of the free layer is consistent with the propagation of a single domain wall in the nanowire after being injected from a reservoir (see Supplementary Information). The hysteresis loop branch that corresponds to the reversal of the free layer can be then explained as follows: (1) injection of the DW at the first GMR electrode at $H_N$, (2) propagation toward the PZT bars (bottom part of the branch), (3) propagation along the PZT bars (centre part of the branch) and then (4) propagation toward the second GMR electrode. Thus the coercive field here corresponds to the average propagation field along the

PZT bar, which is about 6 mT at zero voltage.

The influence of the gate voltages on the domain wall propagation is illustrated in Figure 3a starting from an in-plane depolarised state of the PZT bar. The GMR ratio change with applied voltage is due to the PL magnetisation reversal that occurs before the FL magnetisation is fully reversed. The striking feature seen in Figure 3a is a large increase of the coercivity of the free layer under stress. Following the magnetisation reversal with magnetic field under voltage (see Supplementary Information Fig. S2), we observe first a change of slope of the ascending branch, which is consistent with a modification of the dynamics of DW motion only along the PZT bars. Note also that $H_N$ is not modified under voltage, which is consistent with the injection of the DW outside of the PZT bar. The average propagation field Hp as a function of applied voltage from -50 V to 50 V (100 kV/cm) was extracted at 50% of the total GMR (Figure 3b). It is striking to note that the propagation field of the FL doubles for an applied voltage of $V_{PZT} = \pm 50$ V.

To further confirm that the change of Hp is due to the inverse piezoelectric effect, the propagation field was directly compared with the capacitance of PZT under applied voltages. The Hp was determined by scanning the voltage applied on the PZT layer as shown in the top panel of Figure 3c. We can clearly observe a butterfly shape as expected for a ferroelectric-related effect. The propagation field at 0V is higher than in Figure 3b, which is an indication of a remnant in-plane polarisation of the PZT layer. C-V measurements of the PZT layer were performed at 1 MHz, as seen in the bottom panel of Figure 3c. A butterfly shape is also clearly evidenced with a typical coercivity Ec of about 3 V (6 kV/cm). The observed asymmetrical switching is presumably linked to the asymmetric interface side contact generated from fabrication. The most striking result is that the hysteretic change of Hp with voltage is linked to the switching of the two polarisations of the PZT layer. This clearly demonstrates a magnetoelectric coupling in our hybrid piezoelectric/ferromagnetic nanodevice. Note that any Joule heating can be excluded since the leakage current through the PZT layer is in the nA range, Hp increases with voltage and no increase in the spin-valve resistance is seen with applied voltages.

We propose that our structure, which acts like a magnetic domain wall gate, can form the building block for generating boolean logic functions, as shown in Figure 4a for a NOR gate. The principle is very similar to the concept of magnetic shift register based on moving domain walls with spin-polarised currents[22]. The elementary logic device consists of a domain wall nanowire on top of a PZT bar including two in-puts $B_0$ and $B_1$, one electrical control A and a writing line to generate domain walls. Control A corresponds to injecting a spin polarised current to drive DW motion through spin transfer torque while inputs $B_0$ and $B_1$ allow the application of a gate voltage onto the PZT layer to generate a local energy barrier. The output C is based on a Magnetic Tunnel Junction that can detect the direction of magnetization in the nanowire. First, a DW is created in magnetic wire through an Oersted field generated by a current flowing in a writing line going across the wire. The output C is set to low resistance state '0' (parallel state in the MTJ). When the gate voltage is applied (input $B_0$ is '1' or input B1 is '1' or both are '1'), the DW is pinned, thus output C is '0'. When inputs $B_0$ and $B_1$ are both set to '0', the current drives DW motion toward the PZT electrodes, thus the domain wall propagates through the gate and switches the MTJ (output C is '1'). This scheme corresponds to a NOR gate and it can be reconfigurable as a NAND logic gate by initialising with opposite magnetisation in the wire. Here the two inputs $B_0$ and $B_1$ are exactly equal and can be duplicated along the domain wall nanowire, thus multi-inputs (e.g. 8) logic can be easily realised. We estimate the voltage needed to generate an 80% increase of Hp (as seen in Figure 2) for a typical wire width of 80 nm as follows. Considering an electrical field of 50 kV/cm as reached in this experiment, a voltage of 0.4V is required for the input $B_0$ (or $B_1$), which is the typical value between two resistance states in MTJs. Therefore, the output C can be used as an input for controlling the next logic unit and a complete set of logic functions can be realised based on this simple voltage controlled logic gate, which is promising for low power consumption.

Finally, a direct application of such a domain wall gate concerns the racetrack memory device[22]. Artificial constrictions are required to stabilise domain walls against thermal fluctuations, but further downscaling of such geometries represents a technological challenge. Furthermore, high current densities are required to induce domain wall motion because of the strong pinning as a result of these constrictions. As illustrated in Figure 4b, one solution to circumvent such issues could involve pining domain walls using a voltage-induced strain that acts globally on several domain walls. When domain wall shifts are required, the voltage can be set to zero, thereby reducing the pinning potential and allowing walls to be propagated with low current densities. In this scheme, only the intrinsic domain wall pinning due to defects would contribute to the critical current.

We have demonstrated that domain wall motion in a magnetic nanowire can be electrically controlled at room temperature using strain-mediated magnetoelectric coupling in a piezoelectric/ferromagnetic nanostructure. This is an essential step toward ultra low power spintronics devices. In contrast to recent approaches using electric field effects related to ferromagnetic/oxide interfaces[28-30], the use of piezoelectric materials does not require the ferromagnetic material to be very thin and close to a critical transition to observe a strong effect. In principle, these hybrid ferromagnetic/piezoelectric structures are compatible with any magnetostrictive ferromagnetic material, particularly those having a high thermal stability compatible with the ultimate technological node. Finally, such domain wall gates may be used to pin locally domain walls for racetrack memories and serve as elementary logic gates for spin logic devices.

## Methods Summary

### Fabrication

The devices, which consist of a $PbZr_{0.5}Ti_{0.5}O_3$ film (PZT) with a lateral spin valve grown on top, were fabricated on vicinal $SrTiO_3$ (001) single crystal substrates. The 400 nm PZT was firstly grown by pulsed laser deposition at 580°C under 120 mTorr $O_2$ atmosphere. It was then patterned into a 5 μm wide Hall bar by optical lithography and ion milling using Ar and $O_2$. The 700 nm wide spin valve stripe and the side gate electrodes in the middle part of the bar are fabricated by two steps of electron beam lithography and a subsequent lift-off process. The spin valves consisting 2(nm) MgO/5 CFB/ 4 Cu/ 4 Co/ 0.3 Cu/ 8 IrMn/ 4 Pt were grown by argon sputtering at $5.8 \times 10^{-4}$ mbar Ar under a in plane bias field of 100 mT. The width of 700 nm for the spin valve stripe was chosen to promote easy domain wall propagation by minimising pinning due to rough edges. For the spin valve structure, a 2 nm MgO buffer was used to reduce the electrical field screening effect and domain coupling between the ferroelectric and ferromagnetic layers. A 0.3 nm layer of Cu was inserted between Co and IrMn to enhance the magnetic coupling[31]. The side gate electrodes (5 nm Ti/150 nm Au) were grown by e-beam evaporation and patterned to 4 μm (w) × 400 μm (l). Finally, contact electrodes with 10 nm Ti/ 300 nm Au were deposited using a lift-off process after optical lithography.

**Transport Measurements**

An a.c. (10 µA) lock-in technique was used to obtain the giant magnetoresistance signal at room temperature under an applied voltage on the PZT. The hysteresis loops of GMR were measured by sweeping the magnetic field at a rate of 4 mT/min. Owing to the nature of ferroelectrics, the voltages on PZT were applied in two ways. The first concerns the initial polarising scan, which involves ramping the voltage from 0 V with an in-plane depolarised state to polarised states at high voltages (50 V or -50 V). The second concerns the loop scan, which involves sweeping the voltage from an in-plane polarised state (-50 V) to 50 V and back to -50 V. The leakage currents measured during the voltage scans were measured to be in the nA range.

# Figure Captions

**Figure 1 | Schematic of the lateral approach used to manipulate magnetic domain wall through magnetoelectric coupling. a,** In the absence of applied voltages on the piezoelectric layer, the domain wall (DW) propagates freely in the magnetic stripe. **b,** DW propagation in magnetic stripe can be controlled by voltages through lateral magnetoelectric coupling device. By applying a voltage onto the piezoelectric layer, a local stress is induced, followed by DW blockade. **c,** Measurement configuration with hybrid PZT and spin valve hall bar shaped device, a single DW is injected from a large reservoir. The position of the DW is monitored by measuring the GMR between two electrodes. By applying a voltage on PZT, an induced stress results in a local modification of the domain wall dynamics. The SV multilayer structure is shown on the right.

**Figure 2 | Fabricated device structure and magnetic transport property of SV stripe. a,** Top view of partial fabricated device characterised by optical microscope. The green cross is PZT Hall bar located on the STO substrate, the grey stripe on top of PZT is magnetic SV, and yellow patterns contacted with PZT sides are Ti/Au electrodes, which are well aligned with SV and PZT. **b,** Cross section of the device measured by SEM. SV stripe is located in the middle of PZT, and side electrodes are in a good contact with PZT side walls. c, GMR loops of SV stripe, which is measured by sweeping an external magnetic field H along easy axis. The free layer and pinned layer magnetisation reversal with applied field are shown with solid black squares and open black squares, respectively.

**Figure 3 | Domain wall propagation in the free layer under applied voltages detected with the GMR effect. a,** Giant magnetoresistance loops with different applied voltages, which starts from an unpolarised state of the PZT layer. **b,** Propagation field in the free layer as a function of the voltage applied on the PZT layer. **c,** Propagation field of free layer as measured by scanning the voltage applied on the PZT layer starting from -50V (top panel) and C-V measurement of the PZT layer, starting from -20V (bottom panel).

**Figure 4 | Proposal of domain wall logic and domain wall memory by using piezoelectric/ferromagnetic devices. a,** Design of a multi-inputs NOR logic function by using voltage control of elementary DW gates. Control A is used for current driven DW motion, and the writing line is used to generate a domain wall. Inputs $B_0$ and $B_1$ are piezoelectric controlled DW gates, and output C serves as read out for the magnetization direction in the wire through TMR effect. The corresponding NOR table is shown on top. **b,** Design of a racetrack memory using voltage control of a DW gate. Multiple domain walls can be thermally stabilized when input B is ON whereas current induced domain wall motion through control A can be induced when input B is off.

**Acknowledgements**

We acknowledge invaluable support from V. Pillard, S. Eimer for the film preparation and L. Santandrea for Comsol simulation, and fruitful discussions with E. Fullerton. This work was partially supported by the European FP7 programme through contract NAMASTE number 214499, the European FP7 programme through contract MAGWIRE number 257707 and the ANR-NSF project Friends. N.L. also acknowledges financial support from C'Nano IDF.

**Author contributions**

N.L. designed the experiment, fabricated the device and performed the measurement. N.L. and G.A. did the lithography. N.L. and T.D. analysed and interpreted the data. P.A. and L.D. did the strain simulation. J.V.K. performed analytical and micromagnetic modelling. W.Z. designed logic function. N.L., T.D., L.D., J.-V.K. and D.R. edited and commented on the manuscript. C.C., D.R. and P.L. planned and supervised the project.

**Correspondence**

Correspondence and requests for materials should be addressed to D.R. or N.L. (e-mail: dafine.ravelosona@u-psud.fr; na.lei@u-psud.fr)

Figure 1

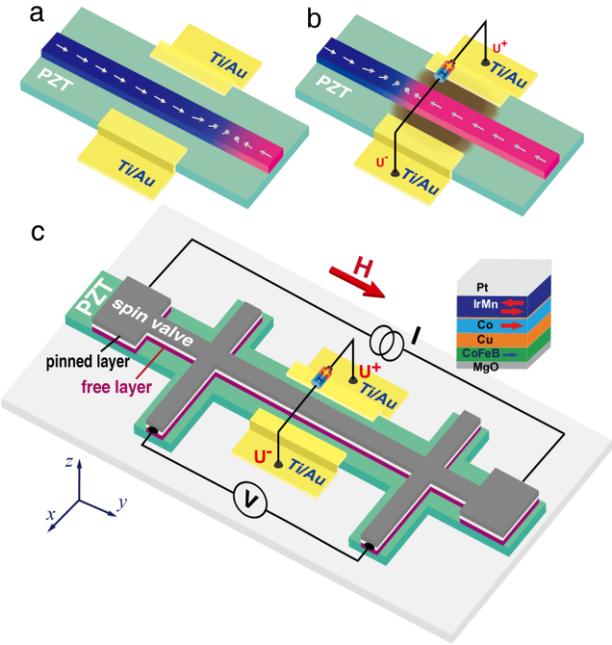

Figure 2

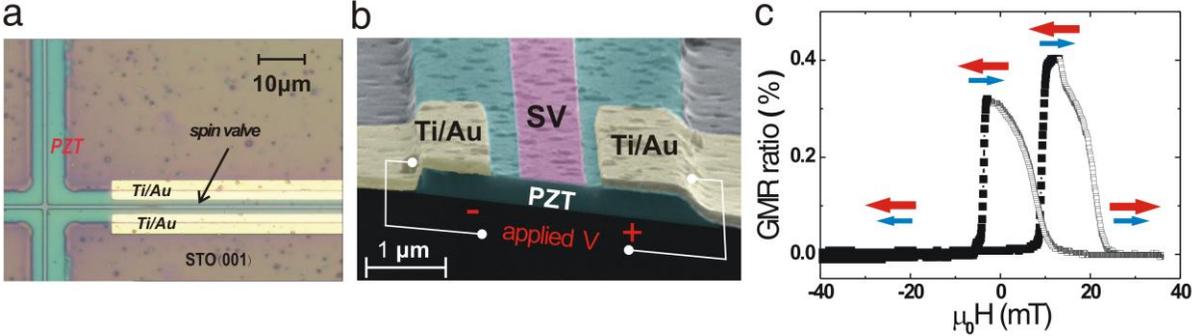

Figure 3

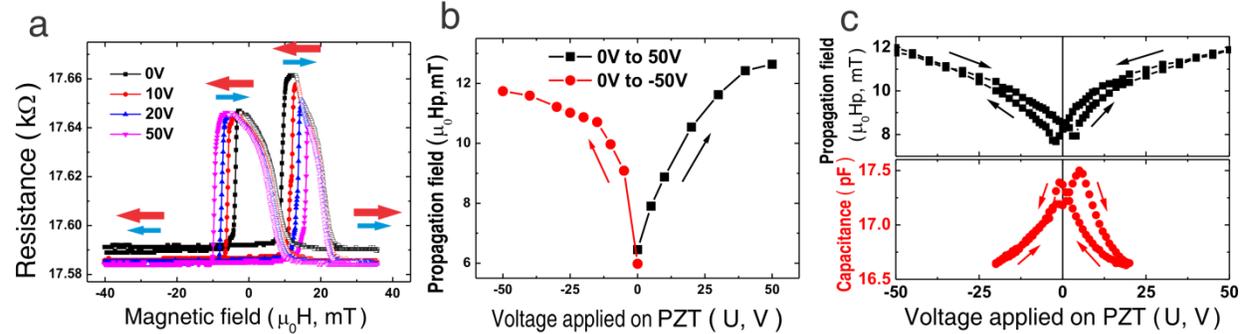

Figure 4

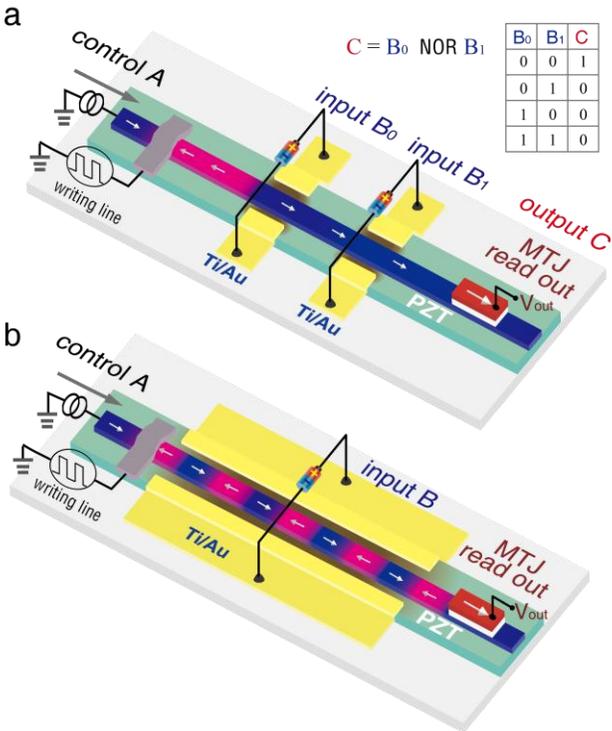